%% file: main.tex
\documentclass[runningheads]{llncs}

\usepackage{a4wide}

\usepackage{xspace}
\usepackage[latin1]{inputenc}
\usepackage{float}
\newfloat{lstfloat}{pthb}{lop}
\floatname{lstfloat}{Listing}

\usepackage{bbding}
\usepackage{graphicx}

\usepackage[svgnames, table]{xcolor}

\usepackage{color, colortbl}
\usepackage{amsmath,amssymb}
\usepackage{booktabs}

\usepackage[hyphens]{url}
\def\ie{{i.e.},~}
\def\eg{{e.g.},~}

\def\z3{{\sc Z3}\xspace}

\colorlet{vert}{green!70!black}
\colorlet{rouge}{red!70!black}
\colorlet{orange}{orange!100!black}
\colorlet{bleu}{cyan!80!white!80!black}
\colorlet{gris}{black!10!white}

\newcounter{mynote}
\newlength\mynotewidth
\makeatletter
\RequirePackage[bookmarks,unicode,colorlinks=true]{hyperref}%
   \def\@citecolor{blue}%
   \def\@urlcolor{blue}%
   \def\@linkcolor{blue}%

\def\orcidID#1{\smash{\href{http://orcid.org/#1}{\protect\raisebox{-1.25pt}{\protect\includegraphics[height=1em]{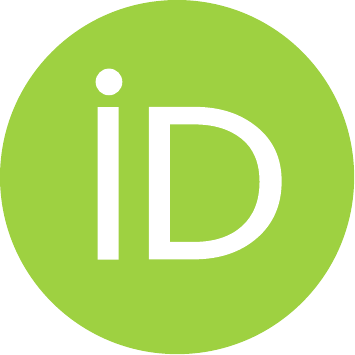}}}}}
\makeatother

\def\dafny{\textsc{Dafny}\xspace}

\usepackage{listings}

\lstdefinelanguage{dafny}{
  sensitive=true,
  keywords={},
  otherkeywords={
  <,>, <==>, <=, >=, |, ==, :=},
  basicstyle=\fontsize{7}{8}\selectfont\ttfamily,
  keywords = [2]{var, trait, datatype,, type, extends, const, address, constructor, predicate, reads, this, map, modifies, function, method, class, lemma, method, ghost, if, then, else, ensures, requires, decreases, while, do, return, od, assert, invariant, int, nat, Msg, seq, tail, init, last, first, take},
  keywordstyle={\bfseries\color{orange}},
  keywordstyle=[2]{\bfseries\color{blue!80!black}},
  identifierstyle=\color{black},
  comment=[l]{//},
  moredelim = [s][\color{gray}\ttfamily]{/**}{*/},
  commentstyle=\color{gray}\ttfamily,
  stringstyle=\color{red}\ttfamily,
  morestring=[b]",
  frame=lines,
  numbers=left,
  xleftmargin=2em
}

\AtBeginDocument{}

\begin{document}

\title{\LARGE \bf Deductive Verification of Smart Contracts \\ with Dafny\thanks{Extended version of the FMICS'22 conference paper.}}
\titlerunning{Deductive Verification of Smart Contracts with Dafny}

\author{
  Franck Cassez (\Envelope)\inst{1,\orcidID{0000-0002-4317-5025}}  \and 
  Joanne Fuller\inst{1} \and 
  Horacio Mijail Ant\'on Quiles\inst{1} 
}

\authorrunning{F. Cassez, J. Fuller and H. M. A. Quiles}

\institute{
    ConsenSys, New York, USA\\
    \email{franck.cassez@consensys.net} \hskip 1em \email{joanne.fuller@consensys.net}
    \hskip 1em \email{horacio.mijail@consensys.net}
  }

\bibliographystyle{splncs04}

\maketitle

\begin{abstract}
We present a methodology to develop verified smart contracts.
We write smart contracts, their specifications and implementations in the verification-friendly language \dafny.
In our methodology the ability to write specifications, implementations and to reason about correctness is a primary concern. 
We propose a simple, concise yet powerful solution to reasoning about contracts that have \emph{external calls}. This includes arbitrary re-entrancy which is a major source of bugs and attacks in smart contracts. 
Although we do not yet have a compiler from \dafny to EVM bytecode,  
the results we obtain on the \dafny code can reasonably be assumed to hold on Solidity code: the translation of the \dafny code to Solidity is straightforward. As a result our approach can readily be used to develop and deploy safer contracts. 
\end{abstract}


\section{Introduction}\label{sec-intro}

\input{intro}

\section{Verification of Closed Smart Contracts}\label{sec-closed}

\input{sc-inv}

\section{Verification Under Adversarial Conditions}\label{sec-revert-extcall}
\input{verif-external}


\section{Conclusion}\label{sec-conclusion}
\input{conclusion}


\end{document}

%% file: intro.tex



The Ethereum network provides the infrastructure to implement a decentralised distributed ledger. 
At the core of the network is the Ethereum Virtual Machine~\cite{yp-22} (EVM) which can execute programs written in EVM \emph{bytecode}.
This remarkable feature means that transactions that update the ledger are not limited to assets' transfers but 
may involve complex business logic that can be executed \emph{programmatically}
by \emph{programs} called \emph{smart contracts}. 

\paragraph{\bf \itshape  Smart Contracts are Critical Systems.}
Smart contracts are programs and may contain bugs.
For example, in some executions, a counter may over/underflow, an array dereference may be outside the range of the indices of the array.  
These runtime errors are vulnerabilities that can be exploited by malicious actors to attack the network: the result is usually a huge loss of assets, being either stolen or locked forever.
There are several examples of smart contract vulnerabilities that have been exploited in the past:
in  2016, a \emph{re-entrance} vulnerability in the Decentralised Autonomous Organisation (DAO) smart contract was exploited to steal more than USD50 Million~\cite{dao-explained}. 
The total value netted from DeFi hacks in the first four months of 2022~\cite{crypto-crumble-22}, \$1.57 billion, has already surpassed the amount netted in all of 2021, \$1.55 billion.


Beyond runtime errors, some bugs may compromise the business logic of a contract: an implementation may contain subtle errors that make it deviate from the initial intended specifications (\eg adding one to a counter instead of subtracting one).

The presence of bugs in smart contracts is exacerbated by the fact that the EVM bytecode of the contract is recorded in the immutable ledger and cannot be updated.
The EVM bytecode of a contract is available in the ledger, and sometimes the corresponding source  code (\eg in Solidity~\cite{solidity}, the most popular language to write smart contracts) is available too, although not stored in the ledger.
Even if the source code is not available, the bytecode can be decompiled  which makes it a target of choice for attackers.
Overall smart contracts have all the features of safety critical systems and this calls for
dedicated techniques and tools to ensure they are reliable and bug-free.

\paragraph{\bf \itshape Smart Contracts are Hard to Verify.}
Ensuring that a smart contract is bug-free and correctly implements a given business logic is hard.
Among the difficulties that software engineers face in the development process of 
 smart contracts are:
\begin{itemize}
    \item The most popular languages Solidity, Vyper~\cite{vyper} (and in the early development stage its offspring Fe~\cite{fe}) to write smart contracts have cumbersome features.
    For instance there is a default \emph{fallback} function that is executed when a contract is called to execute a function that is not in its interface. 
    Some features like the composition of function modifiers have an ambiguous semantics~\cite{DBLP:conf/vstte/Zakrzewski18} and developing a formal semantics of Solidity is still a challenge~\cite{DBLP:conf/sefm/MarmsolerB21}. There are defensive mechanisms (reverting the effects of a transaction, enforce termination with gas consumption) that aim to provide some safety. However, these mechanisms neither prevent runtime errors nor guarantee functional correctness of a contract. 
    \item Most of the languages (\eg Solidity, Vyper for Ethereum) used to develop smart contracts are not \emph{verification-friendly}. It is hard to express safety (and functional correctness properties) within the language itself. Proving properties of a contract usually requires learning another specification language to write specifications  and then embed the source code into this specification language.
    \item Smart contracts operate in an \emph{adversarial environment}. For instance, a contract can call other contracts
    that are untrusted, and that can even call back into the first contract. This can result in subtle vulnerabilities like \emph{re-entrancy}, which are caused by other contracts.    
\end{itemize}

\paragraph{\bf \itshape Our Contribution.}

We present a methodology to develop verified smart contracts.
First, we write smart contracts, their specifications and implementations in the verification-friendly language \dafny.
This is in contrast to most of the verification approaches for smart contracts that build on top of existing languages like Solidity or Vyper and require annotations or translations from one language to another. 
In our methodology the ability to write specifications, implementations and to reason about correctness is a primary concern. 
Second, we use a minimal number of contract-specific primitives: those offered at the EVM level.
This has the advantage of reducing the complexity of compiling a high-level language like \dafny to EVM bytecode.
Third, we propose a simple, concise yet powerful solution to reasoning about contracts that have \emph{external calls}. This includes arbitrary re-entrancy which is a major source of bugs and attacks in smart contracts. 
To summarise, our methodology comprises 3 main steps: 1) reason about the contract in isolation, \emph{closed} contract, Section~\ref{sec-closed}; 2) take into account possible exceptions, Section~\ref{sec-abort}; 3) take into account arbitrary external calls, Section~\ref{sec-external}.
Although we do not yet have a compiler from Dafny to EVM bytecode, 
the results we obtain on the \dafny code can reasonably be assumed to hold on Solidity code: the translation of the \dafny code in Solidity is straightforward. As a result our approach can readily be used to develop and deploy safer contracts. 

\paragraph{\bf \itshape Related Work.}

Due to the critical nature of smart contracts, there is a huge body of work and tools to test or verify them.
Some of the related work targets highly critical contracts, like the deposit smart contracts~\cite{formal-inc-merkle-rv,deposit-cav-2020,DBLP:conf/fm/Cassez21}, including the verification of the EVM bytecode.

More generally there are several techniques and tools\footnote{\url{https://github.com/leonardoalt/ethereum_formal_verification_overview}.} \eg ~\cite{solc-vstte-19,fmbc-20,DBLP:conf/isola/AltR18,DBLP:conf/cpp/AmaniBBS18,csi-mythx,harvey}, for auditing and analysing smart contracts writ\-ten in Solidity~\cite{solidity} or EVM bytecode, but they offer limited capabilities to verify complex functional requirements or do not take into account the possibility of re-entrant calls.

Most of the techniques~\cite{DBLP:conf/blockchain2/HortaRSP20,DBLP:conf/fmcad/DharanikotaMBRL21,DBLP:conf/fm/NehaiB19,DBLP:conf/ccs/BhargavanDFGGKK16,DBLP:series/lncs/SchifflABB20,DBLP:conf/isola/AhrendtB20,DBLP:conf/isola/MarescottiOAEHS20,DBLP:conf/isola/MarescottiOAEHS20,DBLP:conf/vmcai/WesleyCNTWG22}  for the verification of smart contracts using high-level code implement a translation from Solidity (or Michelson for other chains) to some automated provers like Why3, F$^\ast$, or proof assistants like  Isabelle/HOL, Coq.    

The work that is closest to our approach is~\cite{DBLP:journals/pacmpl/BramEMSS21}.
In~\cite{DBLP:journals/pacmpl/BramEMSS21} a principled solution to check smart contracts with re-entrancy is proposed and based on instrumenting the code. Our solution (Section~\ref{sec-external}) is arguably simpler. 
Another difference is that~\cite{DBLP:journals/pacmpl/BramEMSS21} does not use the \emph{gas} resource and is restricted to safety properties. Our approach includes the proof of termination using the fact that each computation has a bounded (though potentially arbitrary large) amount of resources. Modelling the gas consumption is instrumental in the solution we propose in Section~\ref{sec-external} as it enables us to prove termination and to reason by well-founded induction on  contracts with external calls.

%% file: sc-inv.tex

In this section, we introduce our methodology in the ideal case where the code of a smart contract
is \emph{closed}. By closed, we mean that there are no calls to functions outside (\eg an external library or another smart contract) of the contract itself.

\paragraph{\bf \itshape An Abstract View of the EVM.}
The Ethereum platform provides a global computer called the Ethereum Virtual Machine, EVM, to execute smart contracts.

In essence, smart contracts are similar to classes/objects in OO programming languages: they can be created/destructed, 
they have a non-volatile \emph{state}, and they offer some \emph{functions} (interface) to modify their state.
Smart contracts are usually written in high-level languages like Solidity or Vyper and compiled into low-level EVM \emph{bytecode}.
The EVM bytecode of a contract is recorded in the ledger and is immutable.
The state of the contract can be modified by executing some of its functions and 
successive states' changes are recorded in the ledger.

\paragraph{\bf \itshape Transactions and Accounts.}

Participants in the Ethereum network interact by submitting \emph{transactions}.
Transactions can be simple ETH (Ethereum's native cryptocurrency) transfer requests or 
requests to execute some code in a smart contract.
The initiator of a transaction must bound the resources they are willing to use by providing a maximum amount of \emph{gas} $maxG$. In the EVM, each instruction consumes a given (positive) amount of gas.
The execution of a transaction runs until it (normally) ends or until it runs out of gas.
Before running a computation, the initiator agrees on a \emph{gas price}, $gp$, \ie how much one gas unit is worth in ETH.   
At the end of the computation, if there is $gl$ gas units left,\footnote{The EVM tracks the amount of gas left relative to the maximum.} the initiator is charged with
$(maxG - gl) \times gp$ ETH.
To implement this type of bookkeeping, the initiator must have an \emph{account}, the \emph{balance} of which is larger than the maximum fee of $maxG \times gp$ ETH, before executing the transaction.

There are two types of accounts in Ethereum: a \emph{user} account which is associated with a physical owner; and a \emph{contract} account which is associated with a piece of code stored in the ledger. 
Both have a \emph{balance}, stored in the ledger, which is the amount of ETH associated with the account. 
An account is uniquely identified by its (160-bit) \emph{address}. 

\paragraph{\bf \itshape Execution of a Transaction.}


The execution of a transaction involving a contract account can be thought of as a \emph{message call}: an account $m$ sends a message to a contract account $c$ to execute one of its functions $f(\cdot)$ with parameters $x$; this call is denoted  $c.f(x)$. The call can originate from a user account or from a contract account.
When executing $c.f(x)$ some information about the caller $m$ is available such has $m$'s account's address and the maximum amount of gas $m$ is willing to pay for the execution of  $c.f(x)$.
The caller $m$ can also transfer some ETH to $c$ at the beginning of the transaction.
The general form of a transaction initiated by $m$ and involving a contract $c$ is written:
\[
    m \rightarrow  v, g, c.f(x)
\]
where $m$ is the \emph{initiator} of the transaction, $v$ the \emph{amount of ETH} to be transferred to $c$ before executing $f(x)$, and $g$ the \emph{maximum amount of gas} $m$ is willing to pay to execute $c.f(x)$.
To reason about the correctness of smart contracts in a high-level language (not EVM  bytecode), we use some features that are guided by the EVM semantics:
\begin{itemize}
    \item  the values of $m, v, g$ in a transaction are fixed; this means that we can write a transaction as a standard
    method call of the form $c.f(x, msg, g)$  where $msg=(m, v)$ by just adding these values as (read-only) parameters to the original function $f$. We specify all the contracts' functions in this form $c.f(x, msg, g)$.
    In $msg$, $m$ is the \emph{message sender}, $msg.sender$, $v$ the message value, $msg.value$.
    \item The only requirement on the gas consumption is that every function consumes at least one unit of gas, and similar for every iteration of a loop. We use the gas value to reason about termination, and we do not take into account the actual gas cost that only makes sense on the EVM bytecode.
\end{itemize}

\paragraph{\bf \itshape Specification with Dafny.}

To mechanically and formally verify smart contracts, we use the verification-friendly language \dafny~\cite{dafny-ieee-2017}.
\dafny natively supports Hoare style specification in the form of pre- and post-conditions, as well as writing proofs as programs, and offers both imperative, object-oriented and functional programming styles.
The \dafny verification engine checks that the methods satisfy their pre- and post-conditions specifications and also checks for the runtime errors like over/underflows. The result of a verification can be either ``no errors'' -- all the methods satisfy their specifications --, ``possible violation'' of a specification -- this may come with a counter-example -- or the verification can time out.
The form of verification implemented in \dafny is \emph{deductive} as the verifier does not try to synthesise a proof but rather checks that a program adheres to its specification using the available hints. The hints can range from bounds on integer values to more complex lemmas. We refer the reader to~\cite{dafny-ieee-2017} for a more detailed introduction to the language and its implementation. 

\medskip

To model the concepts (transaction, accounts) introduced so far, we provide some data types and an \texttt{Account} trait, Listing~\ref{account}.
A trait is similar to an interface in Java. It can be mixed in a class or in another trait to add some specific capabilities. 
The trait \texttt{Account} provides two members: the balance of the account and its type\footnote{In this paper we do not use any specific features related to the type of an account.} (contract or non-contract which is equivalent to user). 
For example, a user account can be created as an instance of the \texttt{UserAccount} class, line~16. A contract account is created by mixing in the \texttt{Account} trait and by setting the type of the contract accordingly: for instance, the \texttt{Token} contract, Listing~\ref{token}, mixes in \texttt{Account} providing the \texttt{balance} and \texttt{isContract} members. 
For high-level reasoning purposes it is enough to define a type \texttt{Address} as a synonym for \texttt{Account}.

\begin{lstfloat}
\begin{lstlisting}[language=dafny,caption=Datatypes and Account Trait in \dafny., captionpos=t, label={account},numbers=left,xleftmargin=3em]
/** A message. */
datatype Msg = Msg(sender: Account, value: uint256) 

type Address = Account 

/** Provide an Account. */
trait Account {
    /** Balance of the account. */ 
    var balance : uint256

    /** Type of account. */
    const isContract: bool
}

/** A user account. */
class UserAccount extends Account {

    constructor(initialBal: uint256) 
        ensures balance == initialBal
    {
        balance := initialBal;
        isContract := false;
    }
}
\end{lstlisting}
\end{lstfloat}
  
\paragraph{\bfseries \itshape Example: A Simplified Token Contract.}
We now show how to use our methodology to specify, implement and reason about a simple contract: a simplified \texttt{Token} contract.
This contract implements a cryptocurrency: tokens can be minted and transferred between accounts. 
The contract's functionalities are:
\begin{itemize}
    \item the contract's \emph{creator} (an account) can mint new tokens at any time and immediately assign them to an account. This is provided by the \texttt{mint} function;
    \item tokens can be sent from an account \texttt{from} to another \texttt{to} provided the sender's (\texttt{from}) balance allows it. This is provided by the \texttt{transfer} function.
\end{itemize}
The complete \dafny code (specification and implementation) for the \texttt{Token} contract is given in Listing~\ref{token}:
\begin{itemize}
    \item the contract is written as a class and has a \emph{constructor} that initialises the values of the state variables;
    \item each method has a \emph{specification} in the standard form of predicates: the \emph{pre-conditions}, \texttt{requires}, and the \emph{post-condition}, \texttt{ensures}; 
    \item the \texttt{Token} contract has a \emph{global invariant}, \texttt{Ginv()}.
    The global invariant must be maintained by each method call. To ensure that this is the case, \texttt{Ginv()} is added to the pre- and post-conditions of each method\footnote{For the constructor it is only required to hold after the constructor code is executed.} (inductive invariant);
    \item the contract is instrumented with \emph{ghost} variables, and possibly ghost functions and proofs. Ghost members are only used in proofs and do not need to be executable. Moreover, a ghost variable cannot be used to determine the behaviour of non-ghost methods for example in the condition of an \texttt{if} statement;
    \item the $sum(m)$ function is not provided but computes the $sum$ of the \emph{values} in the map $m$;
    \item each method consumes at least one unit of gas and returns the gas left after when it completes. 
\end{itemize}
The \texttt{Token} contract has two non-volatile state variables: \texttt{minter} and \texttt{balances}.
The \texttt{minter} is the creator of the instance of the contract (constructor) and is a \texttt{const}ant, which enforces it can be written to only once.
Initially no tokens have been minted and the map that records the balances (in \texttt{Token}, not ETH) is empty (line~20).
In this specification the creator of the contract is free to deposit some ETH into the contract account.
Note that we can also specify Solidity-like attributes: for instance, \texttt{payable} is a Solidity attribute that can be assigned to a function to allow a contract to receive ETH via a call to this function. If a function is not payable, ETH cannot be deposited in the contract via this function. 
In our setting we can simply add a pre-condition: \texttt{msg.value == 0} (Listing~\ref{token}, line~36).

\begin{lstfloat}
\begin{lstlisting}[language=dafny,caption=A Simple Token Contract in \dafny., captionpos=t, label={token},numbers=left,xleftmargin=3em]
class Token extends Account {

    const minter: Address  //  minter cannot be updated after creation
    var balances : map<Address, uint256>

    ghost var totalAmount: nat  

    /** Contract invariant. */
    predicate GInv() 
        reads this`totalAmount, this`balances
    {
        totalAmount == sum(balances)
    }

    /** Initialise contract.  */
    constructor(msg: Msg) 
        ensures GInv()
        ensures balance == msg.value && minter == msg.sender
    {
        isContract, minter, balances, balance := true, msg.sender, map[], msg.value;
        totalAmount := 0;
    }

    /**
     *  @param  from    Source Address.
     *  @param  to      Target Address.
     *  @param  amount  The amount to be transfered from `from` to `to`.
     *  @param  msg     The `msg` value.
     *  @param  gas     The gas allocated to the execution.
     *  @returns        The gas left after executing the call.
     */
    method transfer(from:Address,to:Address,
        amount:uint256,msg:Msg,gas: nat) returns (g:nat)
        requires from in balances && balances[from] >= amount && msg.sender == from 
        requires gas >= 1
        requires msg.sender == from  && msg.value == 0;
        requires to !in balances ||  
            balances[to] as nat + amount as nat <= MAX_UINT256
        requires GInv()
        ensures GInv()
        ensures from in balances && balances[from] >= old(balances[from]) - amount
        ensures to in balances 
        ensures to != from ==> balances[to] >= amount
        decreases gas
        modifies this
    {
        balance := balance + msg.value;
        var newAmount: uint256 := balances[from] - amount ;
        balances := 
            balances[to := (if to in balances then balances[to] else 0) + amount];
        balances := balances[from := newAmount];
    }  

    /**
     *  @param  to      Target Address.
     *  @param  amount  The amount to receiving the newly minted tokens
     *  @param  msg     The `msg` value.
     *  @param  gas     The gas allocated to the execution.
     *  @returns        The gas left after executing the call.
     */
    method mint(to:Address,amount: uint256,msg:Msg,gas:nat) returns (g:nat)
        requires msg.sender == minter
        requires gas >= 1
        requires to !in balances || 
            balances[to] as nat + amount as nat <= MAX_UINT256
        requires GInv()
        ensures totalAmount == old(totalAmount) + amount as nat
        ensures GInv()
        modifies this`balances, this`totalAmount
    {
        balances := 
            balances[to := (if to in balances then balances[to] else 0) + amount]; 
        //  The total amount increases.
        totalAmount := totalAmount + amount as nat;
        g := gas - 1;
    }
}   
\end{lstlisting}
\end{lstfloat}
    
The global invariant of the contract (line~9) states that the total amount of tokens is assigned to the accounts in \texttt{balances}.
The ghost variable \texttt{totalAmount} keeps track of the number of minted tokens.
The transfer method (line~32) requires that the source account is in the \texttt{balances} map whereas the target account may not be in yet. In the latter case it is added to the map.
Note  that the initiator must be the source account (\texttt{msg.sender == from}, line~36). 

\paragraph{\bfseries \itshape Verification of the Simplified Token Contract.}

The \dafny verification engine can check whether implementations satisfy their pre-/post-conditions.  
In the case of the \texttt{Token} contract, \dafny reports ``no errors'' which means that:
\begin{itemize}
    \item there are no runtime errors in our program. For instance the two requirements \texttt{balances[from] >= amount} (line~34) and  \texttt{balances[to] as nat + amount as nat <= MAX\_UINT256} guarantee that the result of the operation is an \texttt{uint256} and there is no over/underflows at lines~50, 51. 
    The update of a map $m$ is written $m[k := v]$ and results in a map $m'$ such that $m'[w] = m[w], k \neq w$ and $m'[k] = v$ (lines~50, 51, 72).
    \item The global invariant \texttt{GInv()} must be preserved by each method call: if it holds at the beginning of the execution of a method, it also holds at the end. This global invariant must also hold after the constructor has completed. If \dafny confirms \texttt{GInv()} holds everywhere, we can conclude that \texttt{GInv()} holds after any finite number of calls to either \texttt{mint} or \texttt{transfer}.
    \item There are some other pre- and post-conditions that are in the specifications. For example, the \texttt{old} keyword refers to the value of a variable at the beginning of the method and line~41 states that the balance of the \texttt{from} account has been decreased by \texttt{amount}.  
\end{itemize} 

The specification of the \texttt{Token} contract presented in this section assumes the pre-conditions hold for each message (method) call. 
In practice, this has to be ensured at runtime: it is impossible to force an initiator to submit a transaction that satisfies the pre-conditions of a method. 
However, this is a reasonable assumption as in case the pre-conditions do not hold, we can simply abort the execution.
This kind of behaviour is supported by the EVM semantics where it is possible to return 
a status of a computation and 
abort (similar to an exception) the execution of the function and \emph{revert} its effects on the contract's state.
Another more serious simplification of the \texttt{Token} contract is that there is no \emph{external call} to another contract's method.
It turns out that external calls can be problematic in smart contracts and are the source of several attacks.

In the next section we show how to reason about smart contracts under adversarial conditions: exceptions and external calls.


%% file: verif-external.tex
In this section we show how to take into account adversarial conditions: in the first section we describe how to move pre-conditions into runtime checks and enrich our specifications to precisely account for when a function call should revert.
In the second part we propose a general mechanism to capture the possible adversarial effects of external calls.

\subsection{Aborting a Computation}\label{sec-abort}
As mentioned before we cannot enforce the initiator of a transaction to satisfy any pre-conditions when calling a method in a 
smart contract. 
However, a simple way to handle exceptional cases is to explicitly check that some conditions are satisfied 
before executing the actual body of a method, and if it is not the case to abort the computation.
In the EVM semantics this is known as a \emph{revert} operation that restores the state of the contract before the transaction. 
The EVM has a special opcode, \texttt{Revert} to return the status of a failed computation.

In the previous section, we used pre-conditions to write the specification of the methods. 
We can automatically push these pre-conditions into runtime checks at the beginning of each method.
To take into account the possibility of \emph{exceptions} in a clean way, we lift the return values of each method to capture the status of a computation using a standard return generic type of the form \texttt{datatype Try<T> = Revert | Success(v: T)}. If a computation is successful, the value \texttt{v} of type \texttt{T} is returned and boxed in the \texttt{Success} constructor, otherwise \texttt{Revert} is returned.\footnote{\texttt{Revert} is sometimes called \texttt{Failure} and can return a string error message.} 

The implementation of the methods\footnote{The constructor has no pre-condition, so we can assume it always succeeds.}  \texttt{mint} and \texttt{transfer} can be lifted using the return \texttt{datatype Try<T>} as in Listing~\ref{token-revert}, line~1.
This datatype allows for the return of arbitrary values of type \texttt{T} and as a special case when no value is returned, we can set $\texttt{T} = \texttt{Unit}$ the type inhabited by a single value.  

The new code (Listing~\ref{token-revert}) introduces the following features:
\begin{itemize}
    \item the conditions under the first \texttt{if} statements of \texttt{mint} and \texttt{transfer} (respectively at lines~22 and~47) are the negation of the conjunction of all the pre-conditions that are in Listing~\ref{token}.  
    \item The pre-condition \texttt{GInv()} remains in the code. It is not a runtime check but a property of the contract that has to be preserved. This invariant is not part of the executable code. 
    \item In this example of a closed contract we can characterise exactly when the transaction should revert (\texttt{r.T?} is true if and only if \texttt{r} is of type \texttt{T}). For instance the post-condition at line~7 precisely defines the conditions under which the method should not abort.
    \item The post-conditions at lines~15 and~40 ensures that the state of the contract (\texttt{balances}) is unchanged.
\end{itemize}
\dafny returns ``no errors'' for this program, and we can conclude that the global invariant is always satisfied after any number of calls to \texttt{mint} or \texttt{transfer}.
The code for each method does not enforce any pre-condition on the caller and can be translated into runtime checks at the EVM bytecode level. 


\begin{lstlisting}[language=dafny,caption=The Token Contract with Revert., captionpos=t, label={token-revert},numbers=left]
datatype Try<T> = Revert() | Success(v: T) 

method transfer(from:Address,to:Address,amount:uint256,msg:Msg,gas:nat) 
    returns (g: nat, r: Try<()>)

    requires GInv()
    ensures //  if r is of type Success
      r.Success? <==>  
      (from in old(balances) 
      && old(balances[from]) >= amount 
      && msg.sender == from 
      && gas >= 1 
      && (to !in old(balances)||old(balances[to]) as nat + amount as nat<=MAX_UINT256)) 
    /** State is unchanged on an revert. */
    ensures r.Revert? ==> balances == old(balances) 
    ensures g == 0 || g <= gas - 1
    ensures GInv()

    decreases gas
    modifies this
{
    if !(from in balances && balances[from]>=amount && msg.sender==from && gas>=1
        && (to !in balances || balances[to] as nat + amount as nat<=MAX_UINT256) ) {
        return (if gas >= 1 then gas - 1 else 0), Revert(); 
    } 
    var newAmount := balances[from] - amount;
    balances := balances[to := (if to in balances then balances[to] else 0) + amount];
    balances := balances[from := newAmount];
    g, r := gas - 1, Success(());
}  

method mint(to:Address,amount:uint256,msg:Msg,gas:nat) returns (g:nat,r: Try<()>)
    requires GInv()
    ensures r.Success? ==> totalAmount == old(totalAmount) + amount as nat
    ensures r.Revert? <==> 
        !(msg.sender == minter && gas >= 1 && 
            (to !in old(balances)||
                old(balances[to]) as nat + amount as nat<=MAX_UINT256))
    //  state unchanged on a revert.
    ensures r.Revert? ==> balances == old(balances) 
    ensures g == 0 || g <= gas - 1
    ensures GInv()

    modifies this`balances, this`totalAmount
    decreases gas 
{
    if !(msg.sender == minter && gas >= 1 && 
        (to !in balances ||  balances[to] as nat + amount as nat<=MAX_UINT256)) {
        return (if gas >= 1 then gas - 1 else 0), Revert();
    }
    balances := balances[to := (if to in balances then balances[to] else 0) + amount]; 
    //  The total amount increases.
    totalAmount := totalAmount + amount as nat;
    g, r := gas - 1, Success(());
}
\end{lstlisting}

\subsection{Reasoning with Arbitrary External Calls}\label{sec-external}
We now turn our attention to smart contracts that have \emph{external calls}.
The semantics of the EVM imposes the following restrictions on the mutations of state variables for contracts:
the state variables of a contract $c$ can only be updated by a call to a method\footnote{We assume that all methods are \emph{public}.} in $c$.
In other words another contract $c' \neq c$ cannot write the state variables of $c$.

Assume that when we transfer some tokens to a contract via the \texttt{transfer} method, we also \emph{notify} the receiver.
The corresponding new code for \texttt{transfer} is given in Listing~\ref{extern-call}.
If the method \texttt{notify} in contract \texttt{to} does not perform any external call itself, the segment 
\texttt{to.notify($\cdot$)} cannot modify the state variables of the \texttt{Token} contract, and the \texttt{Token} contract invariant \texttt{GInv()} is preserved.
We may not have access to the code of \texttt{notify($\cdot$)} and may be unable to check whether this is the case.

If \texttt{notify} can call another contract it may result in unexpected consequences. For instance if the external call to the method \texttt{to.notify($\cdot$)}
occurs before the update of \texttt{balances[from]}, \texttt{to.notify} may itself call (and collude with) \texttt{from} and call \texttt{from} to do the same transfer again.
As a result many transfers will be performed (as long as some gas is left) and tokens will be \emph{created} without 
a proper call to \texttt{mint}. 
The result is that the number of  minted tokens does not correspond anymore to the number of tokens allocated to accounts, and the global invariant \texttt{Ginv()} does not hold anymore after \texttt{transfer}.
This type of issue is commonly known as the \emph{re-entrancy problem}. 
This vulnerability was exploited in the past in the so-called DAO-exploit~\cite{dao-explained}.

\begin{lstfloat}[t]
\begin{lstlisting}[language=dafny,caption=The Token Contract with a Notification., captionpos=t, label={extern-call},numbers=none]
method transfer(...) returns (g: nat, r: Try<()>)
...
{
    ...
    balances := balances[to := (if to in balances then balances[to] else 0) + amount];
    balances := balances[from := newAmount];
    //  External call to contract `to`. 
    //  If we notify before updating balances, a re-entrant call may drain the contract
    //  of its tokens.
    g, status := to.notify(from, amount, gas - 1);   
    ... 
}  
\end{lstlisting}
\end{lstfloat}

There are several solutions to  mitigate the re-entrancy problem. A simple solution is to require that calls to external contracts occur only as the last instruction in a method (Check-Effect-Interaction pattern~\cite{DBLP:conf/cav/BrittenSR21}).
This is a  \emph{syntactic sufficient condition} to ensure that every update on a contract's state occurs before any external calls. 
This enforces re-entrant calls to happen sequentially.
A \emph{semantic} approach for taking into external calls is proposed in~\cite{DBLP:journals/pacmpl/BramEMSS21} and rely on identifying segments of the code with external calls and adding some local variables to capture the effects of a call and reason about it.

\medskip

We propose a similar but hopefully simpler technique\footnote{It can be implemented directly in \dafny with no need for extra devices.} to model external calls and their effects.
Similar to~\cite{DBLP:journals/pacmpl/BramEMSS21} we do not aim to identify re-entrant calls but we want to \emph{include and model} the effect of possible external (including re-entrant) calls and check whether the contract invariant can be violated or not.
For the sake of simplicity we describe our solution to this problem on the \texttt{to.notify($\cdot$)} example, Listing~\ref{extern-call}, and make the following (EVM enforced) assumptions on \texttt{to.notify($\cdot$)}:
\begin{itemize}
    \item it always terminates and returns the gas left and the status of the call (revert or success),
    \item it consumes at least one unit of gas,
    \item it may itself make arbitrary external calls including callbacks to \texttt{transfer} and \texttt{mint} in the \texttt{Token} contract. As a result there can be complex nested calls to \texttt{transfer} and \texttt{mint}.
\end{itemize}

Our solution abstracts the call to \texttt{to.notify($\cdot$)} into a generic \texttt{externalCall}.
 The new code for the \texttt{transfer} method is given in Listing~\ref{model-extern-call}. 
We model the effect of the external call \texttt{to.notify($\cdot$)} (line~17) by a call to the \texttt{externalCall($\cdot$)} method.

The idea is that \texttt{externalCall($\cdot$)} is going to generate all possible re-entrant calls including nested calls to \texttt{transfer}.
To do so, we introduce some \emph{non-determinism} to allow an external call to callback \texttt{transfer} and \texttt{mint}.
This occurs at lines~51 and~57. Note that the parameters (\texttt{from, to, amount, msg}) of the re-entrant calls are randomly chosen using the \texttt{havoc<T>()} method that returns an arbitrary value of type \texttt{T}.

The code of \texttt{externalCall} works as follows:
\begin{itemize}
    \item non-deterministically pick $k$ and use it to decide whether a re-entrant call occurs or not (lines~42--58). There are three options, and we use $k \mod 3$ to select among them. If $k \mod 3 = 0$ (and there is enough gas left), a re-entrant call to \texttt{transfer} occurs. If  $k \mod 3 = 1$ a re-entrant call to \texttt{mint} occurs. Otherwise, ($k \mod 3 = 2)$, no re-entrant call occurs. 
    \item finally (lines~61--69), we non-deterministically pick a boolean variable $b$ to decide whether a new external call occurs. 
\end{itemize}
We do not provide a formal proof that this captures all the possible re-entrant calls\footnote{This is beyond the scope of this paper.}, but rather illustrate that it models several cases.
First, \texttt{externalCall} can simulate an arbitrary sequence $\texttt{mint}*$ of calls to \texttt{mint}.
This is obtained by selecting successive values of $k$ such that $k \mod 3 = 1$ and then selecting $b = true$.
For instance, the sequence of values $k = 1$, $b = true$, $k = 1$, $b = true$, $k = 2$, $b = false$ simulates two reentrant calls to \texttt{mint}, \ie \texttt{mint}.\texttt{mint}. As the gas value is also a parameter of all the methods and can be arbitrarily large, this model can generate all the sequences of calls in  $\texttt{mint}*$. 
Second, \texttt{externalCall} can also simulate nested \texttt{transfer}/\texttt{mint} calls.
For instance,   the sequence of values $k = 0$, $b = true$, $k = 1$, $b = false$, simulates two reentrant calls to \texttt{transfer} with a nested call to \texttt{mint}.
Third, nested calls to \texttt{transfer} can also be generated by \texttt{externalCall}.
The sequence of values $k = 0$, $b = true$, $k = 0$, $b = false$ simulates two nested re-entrant calls to \texttt{transfer}.

The re-entrant calls can be executed with arbitrary inputs and thus the input parameters are \emph{havoced} \ie non-deterministically chosen and \texttt{externalCall} can generate an arbitrary number of external and re-entrant calls including nested calls to \texttt{transfer} and \texttt{mint}. 


\begin{lstfloat}[hbtp]
\begin{lstlisting}[language=dafny,caption=The Token Contract with External Calls., captionpos=t, label={model-extern-call},numbers=left]
method transfer(from:Address,to:Address,amount:uint256,msg:Msg,gas:nat) 
    returns (g:nat,r:Try<()>)

    ... //  Ensures and requires same as Listing A.3
{ 
    if !(from in balances && balances[from]>=amount && msg.sender==from && gas>=1
        && (to !in balances || balances[to] as nat + amount as nat <= MAX_UINT256) ) {
        return (if gas >= 1 then gas - 1 else 0), Revert(); 
    } 
    var newAmount := balances[from] - amount;
    balances := balances[to := (if to in balances then balances[to] else 0) + amount];
    balances := balances[from := newAmount];
    //  If we swap the line above and the externalCall, 
    //  we cannot prove invariance of GInv()
    //  At this location GInv() must hold which puts a restriction 
    //  on where external call can occur.
    var g1, r1 := externalCall(gas - 1);  // to.notify( from, amount );    
    assert g1 == 0 || g1 <= gas - 1;
    //  We can choose to propagate or not the failure of external call. 
    //  Here choose not to.
    g, r := (if g1 >= 1 then g1 - 1 else 0), Success(());
}  

/**
 *  Simulate an external call with possible re-entrant calls.
 *  
 *  @param  gas The gas allocated to this call.
 *  @returns    The gas left after execution of the call and the status of the call.
 *
 *  @note       The state variables of the contract can only be modified by 
 *              calls to mint and transfer.
 */
 method externalCall(gas: nat) returns (g: nat, r: Try<()>)
    requires GInv()
    ensures GInv()
    ensures g == 0 ||  g <= gas - 1 
    modifies this
    decreases gas 
{
    g := gas; 
    //  Havoc `k` to introduce non-determinism.
    var k: nat := havoc();
    //  Depending on the value of k % 3, 
    //  re-entrant call or not or another external call.
    if k % 3 == 0 && g >= 1 {
        //  re-entrant call to transfer.
        var from: Address := havoc();
        var to: Address := havoc();
        var amount: uint256 := havoc(); 
        var msg: Msg := havoc();
        g, r := transfer(from, to, amount, msg, g - 1);
    } else if k % 3 == 1 && g >= 1 {
        //  re-entrant call to mint. 
        var to: Address := havoc();
        var amount: uint256 := havoc();
        var msg: Msg := havoc();
        g, r := mint(to, amount, msg, g - 1);
    } 
    //  k % 3 == 2, no re-entrant call.
    //  Possible new external call
    var b:bool := havoc();
    if b && g >= 1 {
        //  external call makes an external call. 
        g, r := externalCall(g - 1);
    } else {
        //  external call does not make another external call. 
        g := if gas >= 1 then gas - 1 else 0;
        r := havoc();
    }
}

/** Havoc a given type. */
method {:extern} havoc<T>() returns (a: T)

\end{lstlisting}
\end{lstfloat}

The key ingredient that allows us to reason and prove correctness is the \texttt{gas} value.
We require that \texttt{gas} strictly decreases (line~38) after each recursive call. 
This is stated in \dafny with the \texttt{decreases} clause.
\dafny checks that the value of the \texttt{gas}
parameter strictly decreases on mutually recursive calls. 

Our objective is now to prove, using this model of external calls, that the global invariant \texttt{GInv()} of the contract is always satisfied.
This seems to be a complex task as our model includes an arbitrary and unbounded number of possibly nested external calls.
The result is a mutually recursive program: \texttt{transfer} can call \texttt{externalCall} and  \texttt{externalCall} can call \texttt{transfer} or  \texttt{externalCall}.
However, the property that the \texttt{gas} value strictly decreases on every call enables us to reason by induction.
As the gas decreases on each new call, the induction is well-founded.
And \dafny can indeed prove that the global invariant \texttt{GInv()} is preserved by all the methods including an arbitrary number of possibly re-entrant \texttt{externalCall}s.
Our solution provides a way to model the effects of external calls abstractly but conservatively while still being able to prove properties in modular manner in an adversarial environment modelled by \texttt{externalCall}.
Compared to other approaches we also guarantee termination because we take into account the minimum amount of gas that computations take.

Note that \texttt{externalCall} has the pre-condition \texttt{GInv()}. This means that in \texttt{transfer} the predicate \texttt{GInv()} must be true before the call to \texttt{externalCall}. This amounts to having restrictions on where external calls can occur.
However, without any knowledge of what external calls can do, this seems to be a reasonable restriction. For instance, if the external call has a callback to \texttt{mint} we can only prove the preservation of the invariant \texttt{Ginv()} if it holds before the call to \texttt{mint}.
Of course if we have more information about an external call, \eg we know it does not call back, we can also take it into account with our model: we can adjust \texttt{externalCall} to reflect this knowledge. 

In our example, if we swap the lines~12 and~17 (Listing~\ref{model-extern-call}), \dafny cannot verify that \texttt{GInv()} is preserved by \texttt{transfer}. The reason is that the invariant \texttt{Ginv()} does not hold before the external call.

\medskip

To the best of our knowledge this solution is the first that does not require any specific reasoning device or extension to prove properties of smart contracts under adversarial conditions, but can be encoded directly in a verification-friendly language.

\paragraph{\bfseries \itshape Running Dafny and Reproducing the Verification Results.}

The code used in this paper omits some functions and proofs hints (like \texttt{sum}) and may not be directly verifiable with \dafny.
The interested reader is invited to check out the code in the repository \url{https://github.com/ConsenSys/dafny-sc-fmics} to get 
the full version of our contracts.  
The repository contains the code of the \texttt{Token} contract, a simplified auction contract and
instructions how to reproduce the \dafny verification results. 

The auction contract demonstrates that global invariants (\texttt{GInv()} in \texttt{Token}) are not limited to specifying \emph{state} properties but can also capture two-state or multi-state properties. This can be achieved by adding ghost history variables using sequences.
This type of specifications is expressive enough to encode some standard temporal logic properties on sequences of states of a contract, \eg ``Once the variable ended is set to true, it remains true for ever''  in the simplified auction contract.

In our experiments, \dafny can handle complex specifications and the contracts we have verified are checked with \dafny within seconds on a standard laptop (MacBook Pro). 
The performance does not seem to be an issue at that stage, and if it would become an issue, there are several avenues to mitigate it: \dafny supports modular verification, so we can break down our code into smaller methods; \dafny has built-in strategies to manipulate the verification conditions and break them into simpler ones that can be checked independently.

%% file: conclusion.tex
We have proposed a methodology to model and reason about (Ethereum) smart contracts using the verification-friendly language \dafny.
The main features of our approach are: $i)$ we encode the specifications and implementations of contracts directly in \dafny with no need for any language extensions; $ii)$ we take into account the possibility of \emph{failures} and (arbitrary number of) \emph{external calls}; $iii)$ we specify the main properties of a contract using contract \emph{global invariants}  and prove these properties in a modular manner by a conservative abstraction of external calls with no need to know the code of externally called contracts.

To the best of our knowledge, our abstract model of the effect of external calls is new and the associated proof technique (mutually recursive method calls) is readily supported by \dafny which makes it easy to implement.

We have tested our methodology on several contracts (\eg Token, Simple Auction, Bank) and we believe that this technique can be used to verify larger contracts. 
Indeed, we can take advantage of the modular proof approach (based on pre- and post-conditions) supported by Dafny to design scalable proofs.

Our current work aims to automate the methodology we have presented by automatically generating the different versions of a given contract (closed, revert, external calls) from a simple source contract.   

\medskip

The approach we have presented is general and not exclusive to Dafny, and our methodology can be implemented within other verification-friendly languages like Why3~\cite{DBLP:conf/fm/NehaiB19}, Whiley~\cite{whiley-setss-2018}, or proof assistants like Isabelle/HOL~\cite{Nipkow-Paulson-Wenzel:2002} 
or Coq~\cite{DBLP:conf/laser/Paulin-Mohring11}.   
